\documentclass[10pt,twocolumn]{article}
\usepackage{amsmath}
\usepackage{amssymb}
\usepackage{epsfig}
\usepackage{abstract}
\usepackage{authblk}
\usepackage{color}
\usepackage{wrapfig}
\usepackage[labelfont=bf]{caption}
\usepackage{lipsum}

\renewcommand{\abstractname}{}    

\title{Distributed Hierarchical Control versus an Economic Model for Cloud Resource Management}

\author{Dan C. Marinescu and Ashkan Paya \\
Computer Science Division, EECS Department \\
University of Central Florida, Orlando, Florida 32814, USA \\
Email: [dcm, apaya]@cs.ucf.edu \\
\medskip
John P. Morrison and Philip Healy \\
\medskip
Computer Science Department \\
University College Cork, Cork, Ireland\\
Email: [j.morrison, phealy]@cs.ucc.ie}

\date{\today}

\begin{document}

\maketitle

\begin{abstract}
We investigate a hierarchically organized cloud  infrastructure and compare distributed hierarchical control based on resource monitoring with market  mechanisms for resource management.   The latter do not require a model of the system, incur a low overhead,  are robust, and satisfy several other desiderates of autonomic computing.  We introduce several performance measures and report on simulation studies which show that a straightforward bidding scheme supports an effective admission control mechanism, while reducing the communication complexity  by several orders of magnitude and also increasing the acceptance rate compared to hierarchical control and monitoring mechanisms. Resource management based on market-based mechanisms can be seen as an intermediate step towards cloud self-organization, an ideal  alternative to current  mechanisms for cloud resource management.
\end{abstract}

\section{Introduction}
\label{Introduction}

Cloud computing is a form of utility computing, a service delivery model in which a provider makes computing resources and infrastructure management available to the customer as needed and charges them for the actual resource consumption. Today, utility computing, envisioned by John McCarthy\footnote{In1961, in a speech given to celebrate MIT's centennial, he suggested that in the future computing power and applications could be sold through the utility business model.} and others, is a social and technical reality.

Cloud computing emulates traditional utilities -- such as electricity, water, and gas -- and attempts to reduce costs by assembling large pools of computing resources into data centers. These data centers take advantage of the latest computing and communication technologies, exploit economies of scale, reduce the overhead for delivering services, and provide an appealing computing environment for a large user population. Over the last decade cloud service providers (CSPs) such as Amazon, Google, and Microsoft, have built data centers at unprecedented scales. It is estimated that in 2013 these three CSPs were running roughly one million servers each.

For several decades we have designed and built heterogeneous computing systems with very large numbers of components interacting with each other and with the environment in intricate ways. The complexity of such systems is undeniable but their design was, and still is, based on traditional, mostly deterministic, system organization and management. Ideally this should change, but the path to change is strenuous. 

Resource management in large-scale computing and communication systems such as computer clouds poses significant challenges and generates multiple research opportunities. For example, at this time, the average cloud server utilization is low, while the power consumption of clouds based on over-provisioning is excessive and has a negative ecological impact  \cite{Barroso07, Paya15}. We live in a world of limited resources and cloud over-provisioning is not sustainable either economically or environmentally.

It is generally accepted that distributed systems which maintain state information are neither scalable nor robust; this is the reason why most Internet services are delivered by stateless servers. We have also known for some time that collecting state information consumes a significant share of system resources and that system management decisions based on obsolete state information are far from optimal. This knowledge is critical for the communication and computing infrastructure built around the Internet. In contrast, resource management in cloud computing is still based on hierarchical control models where state information is maintained at several levels. We have also known that assembling large collections of systems each with a small, but finite probability of failure, requires novel design principles to guarantee system availability. 

New strategies, policies and mechanisms to implement these policies are necessary to: allow cloud servers to operate more efficiently; reduce costs for the Cloud Service Providers (CSPs); provide an even more attractive environment for cloud  users; and support some form of interoperability. The pressure to provide new services, better manage cloud resources, and respond to a broader range of application requirements is increasing, as more US government agencies are encouraged to use cloud services\footnote{See for example the November 6 memorandum ``The DoD Cloud Way Forward'' which stresses the need for the DoD to increase its use of cloud services.}.

It is rather difficult given the scale involved to either construct analytical models or to experiment with the systems discussed in this paper and this is precisely the reason why we choose simulation. As stated in \cite{Barroso13}: ``First, they (WSCs) are a new class of large-scale machines driven by a new and rapidly evolving set of workloads. Their size alone makes them difficult to experiment with, or to simulate efficiently.''

The remainder of this paper is organized as follows. In Section \ref{CloudResourceManagement} we analyze the need for alternative strategies for cloud resource management and discuss market based strategies as an intermediate step towards cloud self-organization and self-management. In Section \ref{HierarchicalOrganizationAndControlOfExistingClouds} we examine hierarchical organization and control, which represents the state-of-the-art for the management of existing clouds. To compare hierarchical control prevalent in existing clouds with economic models we conduct a series of simulation experiments and report the results in Sections \ref{SimulationOFHierarchicallyControlledCloudInfrastructure} and \ref{SimulationOfBiddingScheme}, respectively. Finally, in Section \ref{Conclusions} we present our conclusions and discuss future work.

\section{Cloud Resource Management}
\label{CloudResourceManagement}

The policies for cloud resource management can be loosely grouped into five classes: (1) admission control; (2) capacity allocation; (3) load balancing; (4) energy optimization; and  (5) quality of service (QoS) guarantees.

The explicit goal of an {\it admission control} policy is to prevent the system from accepting workload in violation of high-level system policies \cite{Gupta09}. Limiting the workload requires some knowledge of the global state of the system.  {\it Capacity allocation} means to allocate resources for individual instances, where an instance is an activation of a service.   Locating  resources subject to multiple global optimization constraints requires a search in a very large search space when the state of individual systems changes rapidly. 

{\it Load balancing} and {\it energy optimization} are correlated and affect the cost of providing the services; they  can be done locally, but global load balancing and energy optimization policies encounter the same difficulties as the the capacity allocation \cite{Kusic08}. {\it Quality of service} is probably the most challenging aspect of resource management and, at the same time, possibly the most critical for the future of cloud computing.

Resource management policies must be based on disciplined, rather than \textit{ad hoc} methods. The basic mechanisms for the implementation of resource management policies are:

\smallskip

\noindent {\it -1. Control theory.} Control theory uses the feedback to guarantee system stability and to predict transient behavior  \cite{Kusic08}, but can be used only to predict local, rather than global behavior;  applications of control theory to resource allocation are covered in \cite{Dutreilh10}. Kalman filters have been used for unrealistically simplified models as reported in \cite{Kalyvianaki09}, and the placement of application controllers is the topic of \cite{Tang10}.

\smallskip

\noindent {\it -2. Machine learning.}  Machine learning techniques do not need a performance model of the system \cite{Tung07}; this technique could be applied to the coordination of multiple autonomic system managers \cite{Kephart07}.

\smallskip

\noindent {\it -3. Utility-based.} Utility based approaches require a performance model and a mechanism to correlate user-level performance with cost \cite{Kephart11}.

\smallskip

\noindent {\it -4. Economic models.} Economic models such as the one discussed in \cite{Stokely10}, cost-utility models \cite{Bai08},  macroeconomic models \cite{Bai08},   are an intriguing alternative and have been the focus of research in recent years \cite{Buyya09,Marinescu09}.

\medskip

{\bf The need for alternative mechanisms for cloud resource management.} The cloud ecosystem is evolving, becoming more complex by the day.  Some of the transformations expected in the  future add to the complexity of cloud resource management and require different  policies and mechanisms implementing these policies. Some of the factors affecting the complexity of cloud resource management decisions are:

\medskip

\noindent a. {\it Cloud infrastructures are increasingly heterogeneous.}  Servers with different configurations of multi-core processors, attached co-processors (GPUs, FPGAs, MICs), and data flow engines are already, or are expected to become, elements of the cloud computing landscape.  Amazon Web Services (AWS) already support G2-type instances with GPU co-processors.

\medskip

\noindent b. {\it The spectrum of cloud services and cloud applications widens.} For example, in the last year AWS added some 10 new services, including Lambda, Glacier, Redshift, Elastic Cache, and Dynamo DB.  Several types of EC2 (Elastic Cloud Computing) profiles, M3 - balanced, C3 - compute optimized,  R3 - memory optimized, I2 and HS1 - storage optimized were also introduced in recent months. The spectrum of EC2 instance types is also broadening; each instance type provides different sets of computer resources measured by vCPUs (vCPU is a hyper-thread of an Intel Xeon core for M3, C3, R3, HS1, G2, and I2). 

As the cloud user community grows, instead of a discrete menu of services and instance types we expect a continuum spectrum; policies and mechanisms allowing a cloud user to precisely specify the resources it needs and the conditions for running her application should be in place.  At the same time, the cloud infrastructure should  support an increasing number of data- and CPU-intensive  {\it Big Data} applications. 

Many big data applications in computational science and engineering do not perform well on cloud. A 2010 paper \cite{Jackson10}  presents the results of an HPCC (High Performance Computing Challenge) benchmark of {\it EC2} and three supercomputers at NERSC. The results show that the floating performance of EC2 is about 2.5 times lower, 4.6 Gflops versus 10.2  Gflops.  The memory bandwidth is also about 2.5 times lower: 1.7  versus 4.4 GB/sec. The network latency is significantly higher, 145 versus 2.1 $\mu$sec and the network bandwidth is orders of magnitude lower, 0.06 versus 3.4 GB/sec. 

\medskip

\noindent c. {\it Cloud over-provisioning demands high initial costs and leads to low system utilization; this strategy is not economically sustainable \cite{Chang10}.} Cloud {\it elasticity} is now  based on over-provisioning, i.e., assembling pools of resources far  larger than required to satisfy the baseline load.  Elasticity allows cloud users to increase or decrease their resource consumption based on their needs. The average cloud server utilization  is in the 18\% to 30\% range \cite{Barroso07, Barroso13}.  Low server utilization implies that the cloud power consumption is far larger than it should be. The power consumption  of cloud servers is not proportional with the load, even when idle they use a significant fraction of the power consumed at the maximum load. Computers are not {\it energy proportional systems}  \cite{Barroso07} thus, power consumption of clouds based on over-provisioning is  excessive  and has a negative ecological impact. A 2010 survey \cite{Blackburn10} reports that idle or under utilized servers contribute $11$ million tonnes of unnecessary CO$_{2}$ emissions each year and that the total yearly cost for the idle servers is $\$19$ billion.

\medskip

\noindent d. {\it The cloud computing landscape is fragmented.} CSPs support different cloud delivery models: Amazon predominantly IaaS (Infrastructure as a Service), Microsoft PaaS (Platform as a Service), Google mostly SaaS (Software as a Service), and so on. An obvious problem with clear negative implication is vendor lock-in; once becoming familiar and storing her data on one cloud it is very costly for the user to migrate to another CSP.    An organization that can seamlessly support  cloud interoperability and allow multiple  cloud delivery models to coexist poses additional intellectual challenges.

\medskip

{\bf Autonomic computing and self-organization.} In the early 2000s it was recognized that the traditional management of computer systems  is impractical and IBM advanced the concept of {\it autonomic computing} \cite{Ganek03, Kephart03}. However, progress in the  implementation of autonomic computing has been slow.   The  main aspects of autonomic computing  as identified in \cite{Kephart03} are:  {\it Self-configuration} - configuration of components and systems follows high-level policies,  the entire system system adjusts automatically and seamlessly;  {\it Self-optimization }- components continually seek opportunities to improve their own performance and efficiency; {\it Self-healing} - the system automatically detects, diagnoses, and repairs localized software and hardware problems; and  {\it Self-protection }- automatically defend against malicious attacks and anticipate and prevent system-wide failures.  

Autonomic computing is closely related to self-organization and self-management. Several aspects of cloud self-organization are discussed in the literature \cite{Gupta09, Gutierrez10, Lim09}.  Practical implementation of cloud self-organization is challenging for several reasons:

\medskip

\noindent {\it A. The absence of a technically suitable definition of self-organization}, a definition that could hint to practical design principles for self-organizing systems and quantitative evaluation of the results. Marvin Minsky and Murray Gell-Mann \cite{Gell-Mann88} have discussed the limitations of core concepts in complex system theory such as emergence and self-organization.  The same applies to autonomic computing, there is no indication on how to implement any of the four principles and how to measure the effects of their implementation.

\medskip

\noindent {\it B. Computer clouds exhibit the essential  aspects of complexity; it is inherently difficult to control complex systems.}  Complex systems: (a) are nonlinear \footnote{The relation between cause and effect is often unpredictable: small causes could have large effects, and large causes could have small effects. This phenomena is caused by {\it feedback}, the results of an action  or transformation are fed back and affect the system behavior.}; (b) operate far from equilibrium; (c) are intractable at the component level; (d) exhibit different patterns of behavior at different scales; (e) require a long history to draw conclusion about their properties; (f) exhibit complex forms of emergence\footnote{Emergence is generally understood as a property of a system that is not predictable from the properties of individual system components.}; (g) are affected by phase transitions  - for example, a faulty error recovery mechanism in case of a power failure took down Amazon's East Coast Region operations;  and (h) scale well. In contrast, simple systems are linear, operate close to equilibrium, are tractable at component level, exhibit similar patterns of behavior at different levels, relevant properties can be inferred based on a short history, exhibit simple forms of emergence, are not affected by phase transitions, and do not scale well, see also Chapter 10 of \cite{Marinescu13}.  

\noindent {\it C. A quantitative characterization of complex systems and of self-organization is extremely difficult.}  We can only asses the goodness of a particular  self-organization algorithm/protocol indirectly, based on some of the measures of system effectiveness, e.g., the savings in cost or energy consumption. We do not know how far from optimal a particular self-organization algorithm is.

\medskip

{\bf Market-oriented cloud resource management.} The  model we propose uses a market approach based on bidding mechanisms to provide a relatively simple, scalable, and tractable solution to cloud resource allocation, eliminate the need for admission control policies, which require some information about the global state of the system and, most importantly, allow the service to be tailored to the specific privacy, security, and Quality of Service (QoS) needs of each application. At this time, Amazon Web Services (AWS) support  reservations and the so called {\it spot instances}. The reservation system offers instances at a fixed price while the spot instances use a market-based pricing.

The  application of market-oriented policies \cite{Bai08,Marinescu09,Stokely10} and their advantages over the other basic mechanisms implementing resource management policies in large-scale systems have been analyzed in the literature. Control theory \cite{Kalyvianaki09,Kusic08} and utility-based methods require a detailed model of the system and are not scalable.  If no bid exists for a service request then the request cannot be accepted. This procedure acts as an effective admission control. When a bid is generated, the resources are guaranteed to be available. Delivering on that bid is based solely on bidder's local state, so the ability to quantify and to satisfy QoS constraints can be established with a higher degree of assurance. 

Energy optimization decisions such as:  how to locate servers using only green energy, how to ensure that individual servers operate within the boundaries of optimal energy consumption \cite{Barroso07,Paya15}, when and how to switch lightly loaded servers to a sleep state will also be based on local information in an auction-based system. Thus, basing decisions on local information will be accurate and will not require a sophisticated system model nor a large set of parameters that cannot be easily obtained in a practice.

\section{Hierarchical Organization \& Control  of Existing Clouds}
\label{HierarchicalOrganizationAndControlOfExistingClouds}

The large-scale data center infrastructure is based on the so called \textit{warehouse-scale computers} (WSCs)  \cite{Barroso13}. The key insight of the analysis of the cloud architecture in \cite{Barroso13}  is that a discontinuity in the cost of networking equipment leads to a hierarchical, rather than flat, network topology. Furthermore, a limited numbers of uplinks are typically allocated to cross-switch traffic, resulting in disparities between intra- and inter-rack communication bandwidth.

In a typical WSC, racks are populated with low-end servers with a {\tt 1U3} or blade enclosure format. The servers within a rack are interconnected using a local Ethernet switch with 1 or 10 Gbps links. Similar cross-switch connections are used to connect racks to cluster\footnote{The terms \textit{cluster} and \textit{cell} are used interchangeably.}-level networks spanning more than 10,000 individual servers. If the blade enclosure format is being used, then a lower-level local network is present within each blade chassis, where processing blades are connected to a smaller number of networking blades through an I/O bus such as PCIe.

1 Gbps Ethernet switches with up to 48 ports are commodity components. The cost to connect the servers within a single rack -- including switch ports, cables, and server NICs -- costs less than \$30/Gbps per server. However, network switches with higher port counts are needed to connect racks into WSC clusters. These switches are up to 10 times more expensive, per port, than commodity 48 port switches. This cost discontinuity results in WSC networks being organized into a two-level hierarchy, where a limited portion of the bandwidth of rack-level switches (4-8 ports) is used for inter-rack connectivity via cluster-level switches. This inter-rack communications bottleneck can be addressed through the use of higher-capacity interconnects such as Infiniband. Infiniband can scale up to several thousand ports but costs significantly more than even high-end Ethernet switches - approximately \$500-\$2,000 per port. Ethernet fabrics of similar scale are beginning to appear on the market, but these still cost hundreds of dollars per server. Very large scale WSCs can exhaust even the capacity of a two-layer hierarchy with high port-count switches at the cluster level. In this case, a third layer of switches is required to connect clusters of racks.

Using this model, large-scale cloud infrastructures can be viewed as five-level hierarchies with an approximate resource count at each level: $L_{0}$ is composed of individual servers. $L_{1}$ is composed of racks containing $S$ servers connected by a relatively high-speed local network. $L_{2}$ consists of clusters of $R$ racks of lower bandwidth compared to the intra-rack connection bandwidth. $L_{3}$ represents an individual WSC composed of a set of $C$ clusters. At the highest level, $L_{4}$, a cloud infrastructure is composed of $W$ WSCs for a total infrastructure server count of $I = W \cdot C \cdot R \cdot S$.

Analysis of the size and organization of the largest scale cloud infrastructures currently in existence provides insight into real-world values of $W$, $C$, $R$ and $S$. For example, Amazon's largest data center, US East, is located in Virginia, USA. A 2014 analysis used IP ranges to estimate that US East contained 6,382 racks of 64 servers each for a total server count of 408,448 \cite{Liu14}. US East is composed of three availability zones, each of which can be regarded as an independent WSC. Assuming an even distribution of racks between availability zones, we arrive at figure of approximately 136,000 servers per availability zone. However, it is unclear whether the distribution of resources between hierarchical levels corresponds with our model. It is also possible that a flatter or deeper networking hierarchy is used. Nevertheless, we can assume a disparity between intra- and inter-rack connectivity as full-bisection 10 Gbps bandwidth within placement groups is advertised as a selling point for HPC instance types \cite{Amazon2014}.

The disparity between intra- and inter-rack connectivity necessitates that the concept of rack locality is accommodated during the architecture of software systems that are deployed to WSCs. For example, the HDFS filesystem used by Hadoop incorporates rack locality information for redundancy and performance purposes. Swift, the OpenStack object storage service, incorporates locality information so that location-aware services such as HDFS can be run efficiently on top of it. The inter-rack bottleneck also has implications for the management of the infrastructure itself, in particular monitoring.

Given the hierarchical topology of WSC networks and inter-rack bandwidth constraints, a distributed hierarchical management scheme is a natural choice. Several hierarchical cloud management schemes have been discussed in the literature \cite{Addis2013,Andreolini2010,Moens2013}. As the operational details of most large-scale public clouds are not public, the management schemes used by the largest providers -- Amazon, Google, and Microsoft -- can only be surmised.

However, there is at least one large-scale cloud infrastructure where some operational details are available.  Rackspace use OpenStack to manage their public cloud infrastructure and are co-founders of and active contributors to the OpenStack project \cite{Patterson2011}. By default, OpenStack Nova uses the Filter Scheduler, which is centralized \cite{Litvinski2013}. However, Rackspace have contributed to the development of a distributed scheduler, noting in the specification document\footnote{http://wiki.openstack.org/wiki/DistributedScheduler} that \textit{``Rackspace currently uses a distributed model for both scalability and high availability and these same principles need to be brought into Nova to make it suitable for large scale deployments.''}

The distributed scheduler is hierarchical in that it allows the infrastructure to be organized in into trees of nested zones. At the time of writing, Rackspace advertise a total server count of 110,453 \cite{Rackspace2014}, roughly an order of magnitude fewer that those of the largest providers. The fact that Rackspace use a hierarchical management scheme lends support to the idea that the larger providers do also. We therefore postulate that hierarchical management is the \textit{de facto} approach to the management of large-scale cloud infrastructures.

The hierarchical approach to managing WSCs necessitates that service requests arriving at the top level must pass through four levels before they can be satisfied. The decision making processes at each hierarchical level is performed by controllers that oversee the logical unit at each level:

\smallskip

\noindent \textit{WSC Controllers ($L_{3}$):} maintain state information about the $C$ cell beneath it.

\smallskip

\noindent\textit{Cell Controllers ($L_{2}$):} maintain state information on the  $R$ racks beneath it.

\smallskip

\noindent \textit{Rack Controllers ($L_{1}$):} maintain state information on $S$ servers beneath it. 

\smallskip

The hierarchical management of WSCs involves two distinct activities: monitoring and control. Monitoring is the process of gathering the information required for control decisions. Control is the execution of service requests, along with housekeeping activities such as VM migration for power efficiency purposes.

Controllers at Level 3 monitor relevant aspects of server load, such as CPU and memory usage, and communicate this information periodically via messages to the Level 2 controllers. In turn, the controllers at Levels 2 and Level 1 receive state information from the levels below, aggregate it, and communicate this average load information to the level above. The WSC controller maintains the average load of each Level 1 (cell) controller beneath it.

Monitoring messages are sent periodically. Monitoring information is by definition obsolete. The longer the monitoring period the more inaccurate the information is likely to be. However, shortening the monitoring period places more load on the controllers and their networking links.

The hierarchical control model assumes that all service requests arrive first at the WSC controller before being routed through controllers at Levels 1 and 2 before it is fulfilled on a server controller at Level 3. A controller can reject a service request if the logic unit that it represents is overloaded. The optimal case is when none of the controllers a service request is routed through are overloaded and three messages are sent: $L_{3} \rightarrow L_{2}$, $L_{2} \rightarrow L_{1}$ and $L_{1} \rightarrow L_{0}$.

As the load on the system increases, the number of messages required to fulfill a service request ($N_{msg}$) increases as messages are rejected by overloaded controllers and service requests must be rerouted. If there are multiple WSCs in the cloud infrastructure, the worst case scenario is when no WSC can find a cell which can find a rack which can a server fulfill the request: $N_{msg} = N_{M} - 1$ where $N_{M} = W \cdot C \cdot R \cdot S$.

\section{Simulation of a Hierarchically Controlled Cloud Infrastructure}
\label{SimulationOFHierarchicallyControlledCloudInfrastructure}

We  first report on a series of simulation experiments designed to understand the effectiveness of hierarchical control. These experiments were conducted on the Amazon cloud using {\it c3.8xlarge}\footnote{Compute-optimized instance with 32 vCPU and 60 GiB memory.} EC2 instances. It is challenging to simulate systems with 4 to 8 WSCs efficiently, the execution time for each one of the simulation experiments reported in this section is about $24$ hours and each simulation requires 5-6 days wall clock time.

We wanted to understand how the scale and  the load of the system, as well as, several parameters of the resource management system affect the ability of the cloud infrastructure to respond to service requests. An important measure of the hierarchical resource management system effectiveness  is the communication complexity for  monitoring the system and for locating a server capable to process a service request. The communication complexity is expressed by the number of messages at each level of an interconnection infrastructure with different latencies and bandwidth at different levels. 

We simulate hierarchical monitoring and control in a time-slotted system. In each time slot incoming service requests are randomly assigned to one of the WSCs. Each WSC  periodically collects data from the cells, which in turn collect data from racks, which collect data from individual servers. The communication complexity for this monitoring process increases linearly with the size of the system.  The more frequent the monitoring at each level takes place the more accurate the information is, but the larger the volume of data and the interference with the ``productive communication'', communication initiated by running applications.  The communication bandwidth at each level is limited and when the system load increases the communication latency is likely to increase significantly, as many applications typically exchange large volumes of data.

We assume a slotted time; in each reservation slot a batch of requests arrive and the individual requests are randomly assigned to one of the WSCs.  We experiment with two different systems, the first has 4 WSCs and the second 8WSCs. A WSC has the following configuration: 24 cells, 100 racks per cell, 40 servers in each rack, and 4 processors per server. The system is homogeneous, all servers have the same capacity100vCPU. Thus, a WSC has $88,000$ servers and $352,000$ processors. All simulation experiments are conducted for 200 reservation slots and a random batch of service request arrive in each slot. 

Our simulation models a system where load balancers at each level monitor the system they control.  When a request is assigned to a WSC, the load balancer directs it to the cell with the lowest reported load and the process repeats itself at the cell level; the request is directed to the rack with the lowest reported load, which in turn directs it to server with the lowest reported load. If this server rejects the request   the rack load balancer redirects the request to the server with the next lower load. If the rack cannot satisfy the request it informs the cell load balancer which in turn redirects the request to the rack with the next lowest reported average load, and the process ends up with a rejection if none of the cells of the WSC are able to find a server able to satisfy the type, duration, and intensity of the service request.

Our simulation environment is flexible. One first creates a configuration file which describes the system configuration, the network speed and server load and the parameters of the mode, as shown below for the high initial load case.

\begin{verbatim} 
----------------------------------------------------
High initial load simulation 
----------------------------------------------------

% System configuration

 static const int serverNum = 40;
 static const int cpuNum = 4;
 static const int rackNum = 100;
 static const int cellNum = 25;
 static const int WSCsNum = 4;
 static const int servers_capacity =100;

% Network speeds and load parameters  

 static const int interRackSpeed = 1;
 static const int intraRackSpeed = 10;
 static const int MIN_LOAD = 80;
 static const int MAX_LOAD = 85;

% Model parameters
    
 static const int NUMBER_OF_TYPES = 100;
 static const int vCPU_MAX_REQUES = 800;
 static const int vCPU_MIN_REQUEST =10;
 static const int vCPU_PER_SERVER = 10;
 static const int MAX_SERVICE_TIME = 10;
 static const int MONITORING_PERIOD = 10;
 static const int SIMULATION_DURATION = 200;
 static const int TYPES_FOR_SERVER = 5;
 static const int TYPES_FOR_REQUEST = 5;
    
 static const int RACK_CAP = 
                serverNum * servers_capacity;
 static const int CLUS_CAP= 
               rackNum * RACK_CAP;
 static const int WSC_CAP=   
              clusterNum * CLUS_CAP;
 static const int SYSTEM_CAP=   
             WSCsNum * WSC_CAP;
------------------------------------------
\end{verbatim}

The parameters of the simulation experiments have been chosen as realistic as possible; the system configuration is derived from the data in \cite{Barroso13}, the amount of resources in a service request has a broad range, between $10$ and $800$ vCPUs, while a single server can provide $10$ vCPUs. The spectrum of service types offered is quite large, initially $500$ types and then reduced to $100$.  The duration of simulation is limited to $200$ reservation slots by practical considerations regarding costs and time to get the results.

A service request is characterized by three parameters:  (1) The service type;  (2) The service time expressed as a number of time slots; and (3)  The service intensity expressed  as the number of vCPUs needed.

\begin{table*}[!ht]
\caption{Hierarchical control - the simulation results for a system configuration with 4 WSCs. Shown are the initial and final system load for the low and high load, the initial and final coefficient of variation 
$\gamma$ of the load,  the rejection ratio (RR), and the average number of messages for monitoring and control per service request at WSC level, Cell level, and Rack level.}
\label{ResultsHCTab1}
\begin{center}
\begin{tabular} {|c|c|c|c|c|c|c|c|}
\hline
WSCs   & Initial/Final                      & Initial/Final               &       RR       &   \# service          &       WSC          & Cell           &  Rack          \\
 	     &load $(\%)$                     & $\gamma$               &    $(\%)$    &   requests         &    Msg/Req       & Msg/Req   & Msg/Req     \\
\hline
\hline

4	    &22.50/19.78                    &    0.007/0.057          &  2.2           &   14,335,992   &  0.98                &   3.18         &   271.92      \\
            &78.50/82.38                    &    0.004/0.183           &  7.1          &    57,231,592   &  1.01                & 10.16        & 973.15         \\
\hline
\hline 
 8	    &22.50/19.26                    &    0.006/0.049          &  1.9           &    31,505,482   &  0.98                &   3.18         &   271.92      \\
 	    &78.50/81.98                    &    0.005/0.213           &  8.7          &    94,921,663   &  1.01                & 11.36        & 1071.75         \\
\hline
\end{tabular}
\end{center}
\end{table*}

\begin{table*}[!ht]
\caption{Hierarchical control - instead of $500$ different requests types the system supports only $100$; all other  parameters  are identical to the ones of the experiment with the results reported in Table \ref{ResultsHCTab1}.}
\label{ResultsHCTab2}
\begin{center}
\begin{tabular} {|c|c|c|c|c|c|c|c|}
\hline 
   WSCs   & Initial/Final                      & Initial/Final               &       RR       &   \# of service          &       WSC          & Cell           &  Rack      \\
                & load $(\%)$                     & $\gamma$               &    $(\%)$     &   requests                &    Msg/Req       & Msg/Req   & Msg/Req  \\
\hline 
\hline
    4         & 22.50/21.15                    &   0.003/0.051           &  1.9             &     16,932,473           &  1.00               &   3.53        &   337.34    \\
               & 82.50/67.18                    &   0.003/0.109           &   7.2            &     42,034,225           & 1.00                &  11.15       & 1,097.00    \\
 \hline 
 \hline
    8         & 22.50/22.13                   &    0.008/0.055           &  5.4             &     38,949,889         &   1.00               &   4.22       &    470.35   \\
               & 82.50/81.63                   &   0.006/0.155            &  4.2             &     84,914,877        &  1.00               & 10.72       &  1,038.96     \\
\hline
\end{tabular}
\end{center}
\end{table*}

\begin{table*}[!ht]
\caption{Hierarchical control - instead of $5$ different service types a server offers only $2$; all other  parameters  are identical to the ones of the experiment with the results reported in Table \ref{ResultsHCTab1}.}
\label{ResultsHCTab3}
\begin{center}
\begin{tabular} {|c|c|c|c|c|c|c|c|}
\hline 
   WSCs   & Initial/Final                      & Initial/Final               &       RR       &   \# of service          &       WSC        & Cell           &  Rack      \\
                & load $(\%)$                     & $\gamma$               &    $(\%)$     &   requests                &    Msg/Req     & Msg/Req   & Msg/Req  \\
\hline 
\hline
    4         & 22.50/21.15                    &   0.003/0.051           &     1.7          &      17,341,885          &  0.99              &    3.22      &     276.34   \\
               & 82.50/74.27                    &   0.006/0.059           &   14.6           &    52,206,014         &  1.00               &  12.12      &   1255.40  \\
 \hline 
 \hline
    8         & 22.50/16.27                   &    0.006/0.035           &     1.3          &       37,750,971         &   0.99               &   3.18       &    268.27  \\
               & 82.50/74.55                   &    0.007/0.081           &     2.9          &     99,686,943         &  1.00                & 10.77       & 1,036.64  \\
\hline
\end{tabular}
\end{center}
\end{table*}

\begin{table*}[!ht]
\caption{Hierarchical control - the service time is uniformly distributed in the  range $(1- 20)$ reservation slots; all other parameters are  identical to the ones of the experiment with the results reported in Table \ref{ResultsHCTab1}.}
\label{ResultsHCTab4}
\begin{center}
\begin{tabular} {|c|c|c|c|c|c|c|c|}
\hline 
   WSCs   & Initial/Final                      & Initial/Final               &       RR       &   \# of service          &       WSC        & Cell           &  Rack      \\
                & load $(\%)$                     & $\gamma$               &    $(\%)$     &   requests               &    Msg/Req     & Msg/Req   & Msg/Req  \\
\hline 
\hline
    4         & 22.50/22.41                    &   0.005/0.047            &     0.20         &      12,352,852        &     1.00            &  3.13        &    261.11   \\
               & 82.50/80.28                    &   0.003/0.063            &     2.10         &      43,332,119      &     1.00            &  3.41       &    1108.12  \\
 \hline 
 \hline
    8         & 22.50/22.77                   &    0.005/0.083           &     1.30           &       25,723,112       &   1.00             &    3.11      &    236.30   \\
               & 82.50/79.90                   &    0.005/0.134           &     4.10           &     88,224,546        &  1.00             &  10.63      &  1029.56   \\
 \hline
\end{tabular}
\end{center}
\end{table*}

\begin{table*}[!ht]
\caption{Hierarchical control - the monitoring interval is increased from $10$ to $50 $ reservation  slots all other parameters  are identical to the ones of the experiment with the results reported in Table \ref{ResultsHCTab1}.}
\label{ResultsHCTab5} 
\begin{center}
\begin{tabular} {|c|c|c|c|c|c|c|c|}
\hline 
   WSCs   & Initial/Final                   & Initial/Final               &       RR       &   \# of service          &       WSC        & Cell           &  Rack      \\
                & load $(\%)$                  & $\gamma$               &    $(\%)$     &   requests               &    Msg/Req     & Msg/Req   & Msg/Req  \\
\hline 
\hline
    4         & 22.50/21.07                 &   0.003/0.033            &     1.00         &      12,335,103       &  0.99              &    3.21      &     270.07   \\
               & 82.50/83.46                 &   0.007/0.080            &     1.80         &    51,324,147         &  1.01              &  10.87      &   1040.63   \\
 \hline 
 \hline
    8         & 22.50/19.16                &    0.005/0.030            &     1.30         &       29,246,155      &   1.00             &    3.37    &    304.88  \\
               & 82.50/84.12                &    0.002/0.041            &      2.30        &       93,316,503      &   1.00             &    3.66    &    1005.87 \\
\hline
\end{tabular}
\end{center}
\end{table*}

\begin{figure*}[!ht]
\begin{center}
\includegraphics[width=8.7cm]{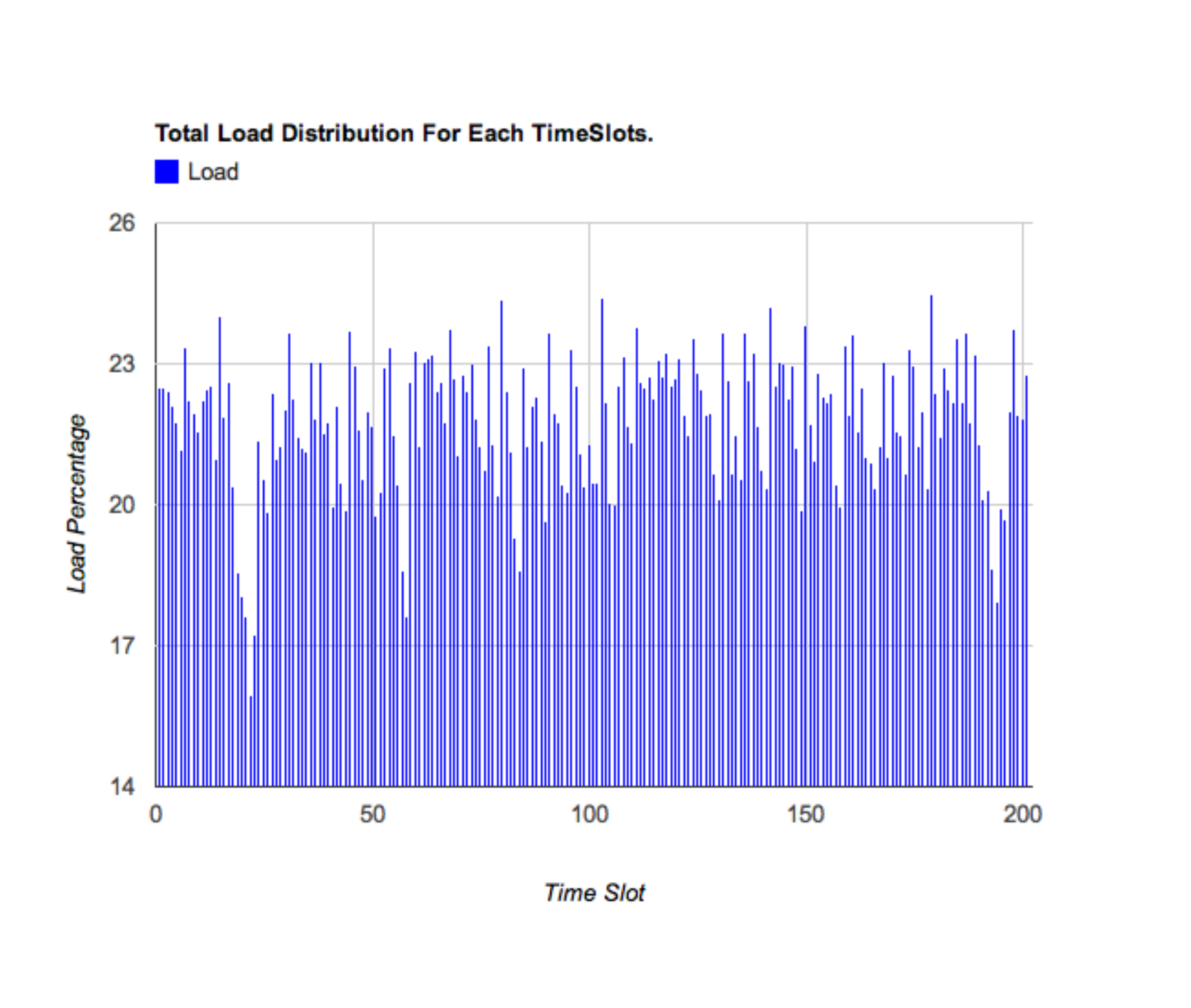}
\includegraphics[width=8.7cm]{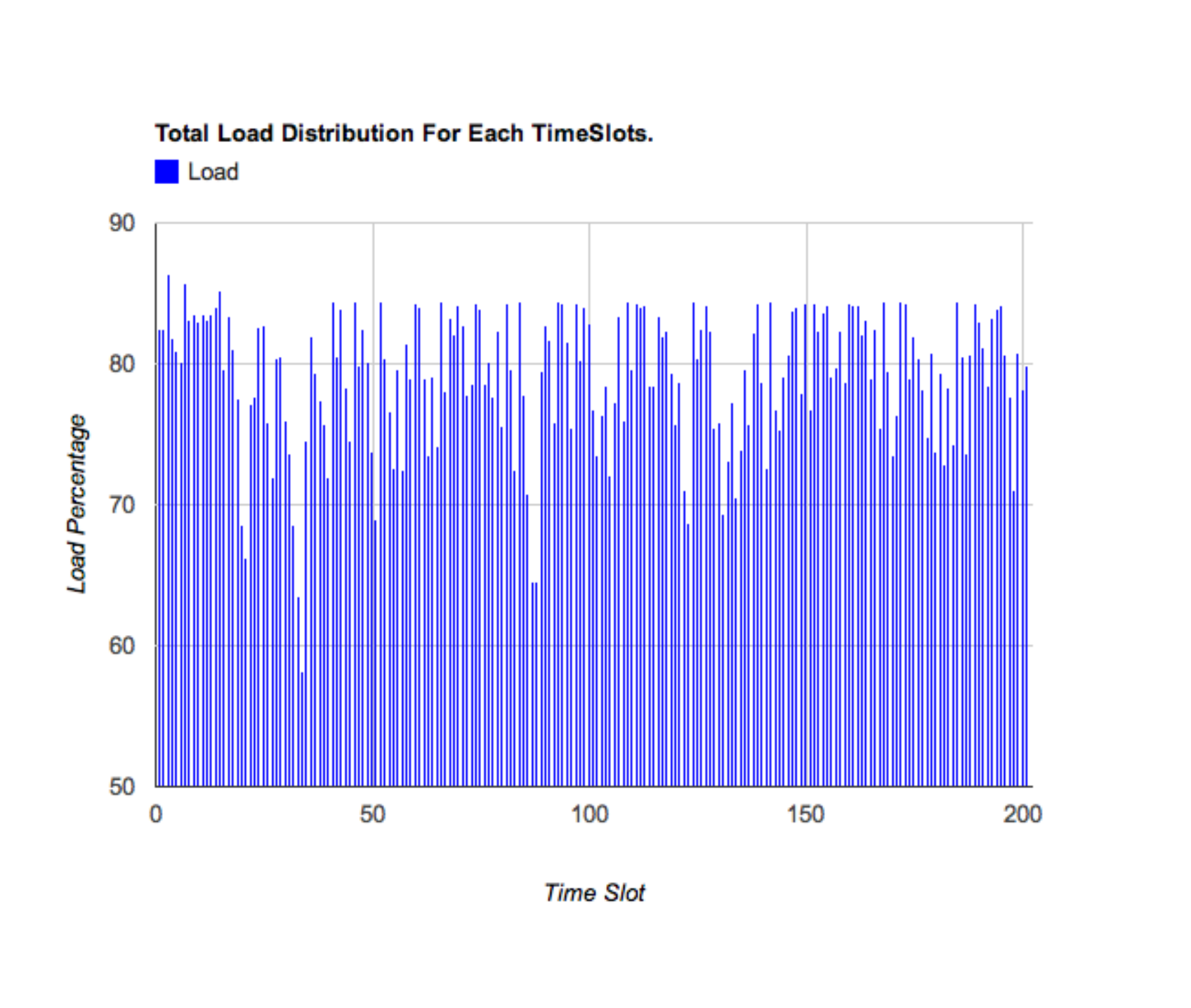}\\
\end{center}
\caption{Hierarchical control - time series of the average load  of a cloud with eight WSCs. The monitoring interval is $20$ reservation slots and the service time is uniformly distributed in the range $1 - 20$ reservation slots. The initial average system load is: (Left) $20\%$; (Right) $80\%$ of system capacity.}
\label{ResultsHC-Fig1}
\end{figure*}

\begin{figure*}[!ht]
\begin{center}
\includegraphics[width=8.7cm]{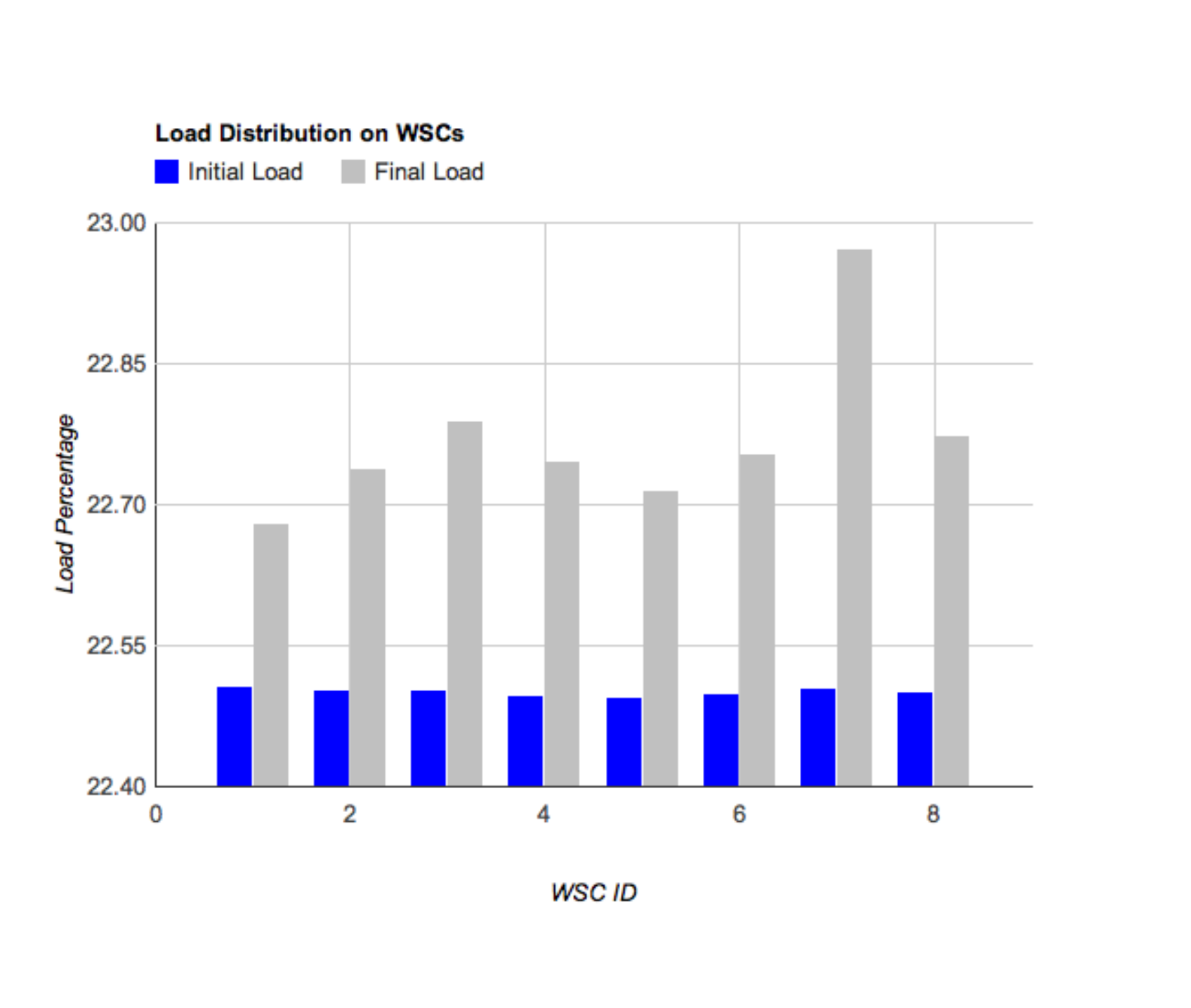}
\includegraphics[width=8.7cm]{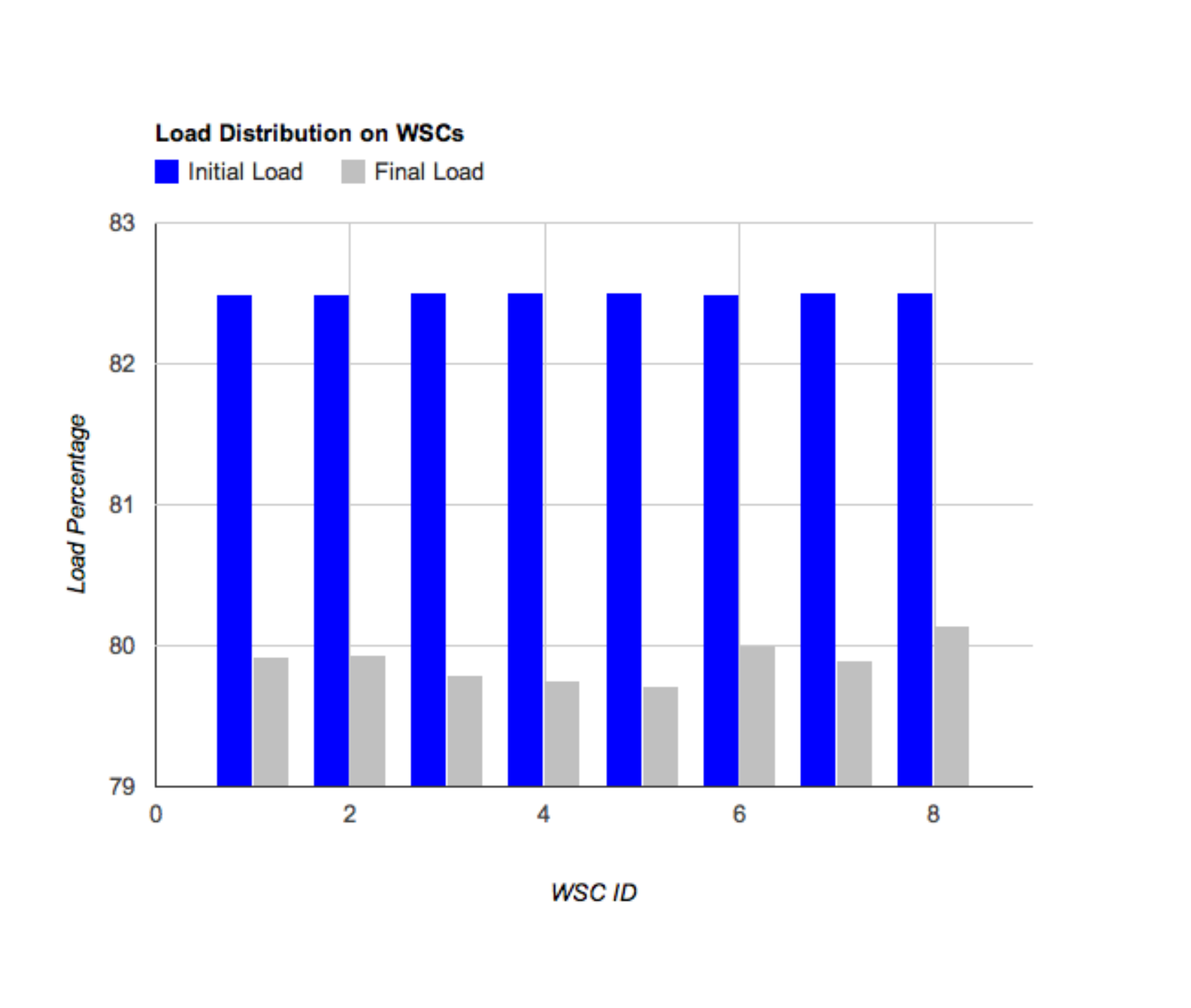}\\
\end{center}
\caption{Hierarchical control - initial and final average load of a cloud with eight WSCs. The monitoring interval is $20$ reservation slots and the service time is uniformly distributed in the range $1 - 20$ reservation slots. The initial average system load is: (Left) $20\%$; (Right) $80\%$ of system capacity.}
\label{ResultsHC-Fig2}
\end{figure*}

We study the impact of  the system size, the system load, the monitoring interval, the number of service types, the  number of service types supported by a server and of the service time on important systems parameters such as:

\smallskip

\noindent - a. The number of messages exchanged at different levels for mapping the service requests. These numbers reflect the overhead of the request processing process.

\smallskip

\noindent -b. The ability of the system to balance the load measured as the coefficient of variation (the variance versus the average) of the system load per reservation slot.

\smallskip

\noindent -c. The rejection ratio, the ratio of service requests rejected because there could be found no server able to match the service type, the service intensity, and the service duration demanded by the clients.

\smallskip

The simulation is conducted for two average initial system loads: low, around $20\%$ and high, around $80\%$ of the system's capacity. The total number of service requests for 4 WSCs and for low and high initial system load are around $(12 -17) \times 10^{6}$ and $(42 - 57) \times 10^{6}$, respectively.  In each case we show the number of WSCs, the initial and final system load for the low and high load, the initial and final coefficient of variation $\gamma$ of the load,  the rejection ratio (RR), and the number of messages for monitoring and control per service request at WSC level, Cell level, and Rack level.

For the first experiment the attributes of service requests are uniformly distributed and the ranges are:  $(1 - 100)$,   $(1 - 10)$, and  $(10-800)$ for service type, service time, and service intensity, respectively.  A server supports $5$ different service types randomly selected from a total of $500$ possible service types. The monitoring interval is $10$ reservation slots; for later experiments it will increase to $20$ and then to $50$ reservation slots. 

The results of our first simulation experiment in Table \ref{ResultsHCTab1} show that the rejection ratio, the coefficient of the variation of the final load, and the average number of messages required to map a service request to a server are more than three fold larger in the case of higher load; indeed, $7.1/2.2=3.22, 0.183/0.057 = 3.22$, and $984/276=3.2$. At higher load more requests are rejected, load balancing is less effective, and the overhead for mapping a request is considerably higher. The increase in the number of messages means a substantial increase of the communication costs and also a longer waiting time before a request enters the service. 

Doubling the size of the system does not affect the statistics for the same average system load. For example, when the initial average load is $22.50\%$ the average number of messages exchanged per service request is the same at the three levels of the hierarchy for both system configurations. The rejection ratio varies little,  $2.2$ versus $1.9\%$ and $7.1\%$ versus $8.7\%$ for $4$ and $8$ WSCs, respectively.

\smallskip

Next we explore the effects of changing various parameters of the system model. Our experiments investigate the effects of: (1) Doubling the number of WSCs from 4 to 8; (2)  Reducing the number of types of services from $500$ to $100$; (3) Reducing the number of types of services offered by each server from $5$ to $2$; and (4) Increasing the monitoring interval to 50 time slots and changing the distribution of the service time; initially it was uniformly distributed in the interval $(1-10)$ time slots and this interval will be $(1-20)$ time slots.

Table \ref{ResultsHCTab2} presents the results after reducing the total number of service request types from $500$ to $100$. We see a reduction of the rejection ratio and of the number of messages  at high load for  the larger configuration of 8 WSCs compared to the case in Table  \ref{ResultsHCTab1}. We also notice that in this case the rejection ratio decreases from $7.4\%$ to $4.2\%$ when we increase the size of the system from $4$ to $8$ WSCs.

Table \ref{ResultsHCTab3} presents the results when the number of service types offered by a  server is reduced  from $5$ to just $2$.  We see again a reduction of the rejection ratio at high load when we double the size of the system. The drastic reduction of this ratio, from 14.6 to 2.9 can be attributed to the fact that  we randomly assign an incoming service request to one of the WSCs and the larger the number of WSCs the less likely is for the request to be rejected.  The number of messages at the rack level is considerably larger for the smaller system configuration at high load, 1255 versus 973 in the first case presented in Table \ref{ResultsHCTab1}. 

Next we set the monitoring interval to $20$ reservation slots and the service time is now uniformly distributed in the range $1- 20$ reservation slots. The results  in Table \ref{ResultsHCTab4} show that the only noticeable effect is the reduction of the rejection rate compared with the case in Table \ref{ResultsHCTab1}. In the following experiment we extended the monitoring interval from  $10$ to  $50$ reservation slots. Recall that the service time is uniformly distributed in the range $1$ to $10$ reservation slots; even when the monitoring interval was $10$ reservation slots, this interval is longer than the average service time thus,  the information available to controllers at different levels is obsolete. The results in Table \ref{ResultsHCTab5} show that  increasing the monitoring interval to $50$ slots has little effect for the 4 WSC configuration at low load,  but it reduces substantially the rejection ratio and increases the number of messages at high load. For the 8 WSC configuration increasing the monitoring interval reduces the rejection ratio at both low and high load, while the number of messages changes only slightly.

Figures \ref{ResultsHC-Fig1} and  \ref{ResultsHC-Fig2} refer to the case presented in Table \ref{ResultsHCTab1}  when the monitoring interval is $20$ time slots and the service time is uniformly distributed in the $1 - 20$ slots range and there are 8 WSCs.  Figures \ref{ResultsHC-Fig1}(Left) and (Right) show the time series of the average system load for the low and the high initial load, respectively.  We see that the actual system workload has significant variations from slot to slot; for example, at high load the range of the average system load is from $58\%$ to $85\%$ of the system capacity.  \ref{ResultsHC-Fig2} show the initial and the final load distribution for the 8 WSCs; the imbalance among  WSCs at the end of the simulation is in the range of $1-2\%$. 

The results of the five simulation experiments are consistent, they typically show that at high load the number of messages, thus the overhead for request mapping increases three to four fold,  at both cell and rack level and for both system configurations, 4 and 8 WSCs.

\section{Simulation of an Economic Model of Cloud Resource Management}
\label{SimulationOfBiddingScheme}

In this section we discuss the simulation of an economic model for cloud resource management based on a straightforward bidding scheme.  There is no monitoring and in each reservation slot all servers of a WSC bid for service. A bid consists of the service type(s) offered and the available capacity of the bidder. The overhead is considerably lower than that of the hierarchical control; there is no monitoring and the only information maintained by each WSC consists only of the set of unsatisfied bids  at any given time. The servers are autonomous but act individually, there is no collaboration among them while self-organization and self-management require agents to collaborate with each other. 

At the beginning of a reservation slot servers  with available capacity at least as large as a given threshold $\tau$ place bids which are then collected by each WSC. A bid is persistent, if is not successful in the current reservation slot it remains in effect until a match with a service request is found. This strategy to reduce the communication costs is justified because successful bidding is the only way a server can increase its workload.  

We investigate the effectiveness of this mechanism for lightly loaded, around $20\%$ average system load, and for heavily loaded, around $80\%$ average  system load. The thresholds for the two cases are different, $\tau=30\%$ for the former and $\tau=15\%$ for the latter. The choice for the lightly loaded case is motivated by the desire to minimize the number of messages; a large value of $\tau$, e.g., $40\%$ would lower the rejection ratio but increase the number of messages. In the heavily loaded system case increasing the threshold, e.g., using a value $\tau=20\%$, would increase dramatically the rejection rate; indeed, few servers would have $20\%$ available capacity when the average system load is $80\%$.

\begin{table*}[!ht]
\caption{Market-based mechanism - simulation results for a system configuration with 4 WSCs. Shown are the initial and final system load for the low and high load, the initial and final coefficient of variation 
$\gamma$ of the load,  the rejection ratio (RR), and the average number of messages for monitoring and control per service request at WSC level, Cell level, and Rack level.}
\label{ResultsBidTab1}
\begin{center}
\begin{tabular} {|c|c|c|c|c|c|c|c|}
\hline
WSCs   & Initial/Final                      & Initial/Final               &       RR       &   \# service          &       WSC          & Cell           &  Rack          \\
 	     &load $(\%)$                     & $\gamma$               &    $(\%)$    &   requests         &    Msg/Req       & Msg/Req   & Msg/Req     \\
\hline
\hline
4	    &22.50/23.76                    &    0.007/0.067          &  .22           &     15,235,231   &  0.002                &   0.011         &   0.987      \\
            &82.50/80.32                    &    0.004/0.115           &  5.44         &     63,774,913  &  0.003                &   0.042         &   4.155         \\
\hline
\hline 
 8	    &22.50/22.47                    &    0.006/0.033          &  .18           &     30,840,890   &  0.002                &   0.011         &   0.987      \\
 	    &82.50/81.30                    &    0.005/0.154          &  7.23         &     89,314,886   &  0.003                &   0.054         &   5.761         \\
\hline
\end{tabular}
\end{center}
\end{table*}

\begin{table*}[!ht]
\caption{Market-based mechanism -  instead of $500$ different requests types the system supports only $100$; all other parameters are  identical to the ones of the experiment with results reported in Table \ref{ResultsBidTab1}.}
\label{ResultsBidTab2}
\begin{center}
\begin{tabular} {|c|c|c|c|c|c|c|c|}
\hline 
   WSCs   & Initial/Final                      & Initial/Final               &       RR       &   \# of service          &       WSC          & Cell           &  Rack      \\
                & load $(\%)$                     & $\gamma$               &    $(\%)$     &   requests                &    Msg/Req       & Msg/Req   & Msg/Req  \\
\hline 
\hline
    4         & 22.50/22.3                    &   0.004/0.050           &  .18             &     15,442,372           &  0.002            &   0.011       & 0.987    \\
               & 82.50/79.88                  &   0.004/0.098           &   6.01          &     56,704,224           &  0.002            &   0.059       & 5.968    \\
 \hline 
 \hline
    8         & 22.50/23.0                   &    0.007/0.049           &  .3             &     31,091,427         &   0.002               &   0.011       &    0.987   \\
               & 82.50/80.91                 &    0.009/0.127           &  5.81         &     85,322,714         &   0.003               &   0.0.51      &    5.845     \\
\hline
\end{tabular}
\end{center}
\end{table*}

\begin{table*}[!ht]
\caption{Market-based mechanism - instead $5$ different service types a server offers only $2$; all other parameters are  identical to the ones for the experiment with results reported in Table \ref{ResultsBidTab1}.}
\label{ResultsBidTab3}
\begin{center}
\begin{tabular} {|c|c|c|c|c|c|c|c|}
\hline 
   WSCs   & Initial/Final                      & Initial/Final               &       RR       &   \# of service          &       WSC        & Cell           &  Rack       \\
                & load $(\%)$                     & $\gamma$               &    $(\%)$     &   requests               &    Msg/Req     & Msg/Req   & Msg/Req  \\
\hline 
\hline
    4         & 22.50/20.94                    &   0.007/0.056           &     .1          &     15,295,245          &  0.002              &    0.011      &     0.987   \\
               & 82.50/77.83                    &   0.008/0.133           &     10.1      &     49,711,936          &  0.003              &    0.063      &     6.734  \\
 \hline 
 \hline
    8         & 22.50/22.33                   &    0.007/0.063           &     .02          &       31,089,191         &   0.002           &   0.011       &    0.987  \\
               & 82.50/78.18                   &    0.008/0.142           &     3.61        &       71,873,449         &  0.002            &   0.059       &    6.098 \\
\hline
\end{tabular}
\end{center}
\end{table*}

\begin{table*}[!ht]
\caption{Market-based mechanism -  the service time is uniformly distributed in the range $(1 - 20)$ reservation slots; all other parameters are  identical to the ones of the experiment with  results reported in Table \ref{ResultsBidTab1}.}
\label{ResultsBidTab4} 
\begin{center}
\begin{tabular} {|c|c|c|c|c|c|c|c|}
\hline 
   WSCs   & Initial/Final                      & Initial/Final               &       RR       &   \# of service          &       WSC        & Cell           &  Rack      \\
                & load $(\%)$                     & $\gamma$               &    $(\%)$     &   requests               &    Msg/Req     & Msg/Req   & Msg/Req  \\
\hline 
\hline
    4         & 22.50/23.31                    &   0.002/0.064            &     2.27         &      13,445,186         &  0.001           &    0.011      &     0.988   \\
               & 82.50/84.05                    &   0.007/0.101            &     3.75         &      57,047,343         &  0.002           &    0.042     &     6.329   \\
 \hline 
 \hline
    8         & 22.50/18.93                   &    0.007/0.038            &     2.94           &       28,677,012       &   0.001          &    0.011    &    0.988  \\
               & 82.50/85.13                   &    0.008/0.072            &     4.38          &      88,342,122      &   0.002          &    0.029    &    4.078 \\
\hline
\end{tabular}
\end{center}
\end{table*}

To provide a fair comparison we repeat the measurements reported in Section \ref{SimulationOFHierarchicallyControlledCloudInfrastructure} under the same conditions but with bidding replacing the monitoring and hierarchical control. We use the same performance indicators as the ones reported in Section \ref{SimulationOFHierarchicallyControlledCloudInfrastructure},  those measuring communication complexity, the efficiency of load balancing, and the rejection ratio. The results are shown in Tables  \ref{ResultsBidTab1} - \ref{ResultsBidTab4}.

We notice a significant reduction of the communication complexity, more than two orders of 
magnitude in case of the market-oriented mechanism. For example,  at low average load the average number of messages per reservation request at the rack level  is $0.987$ (see Table \ref{ResultsBidTab1})  versus $271.92$ reported in Table \ref{ResultsHCTab1} for the 4  and for the 8 WSCs case.   At high average load the same figures are: $4.155$  versus $973.14$ for the $4$ WSC case and $5.761$ versus $1071.75$ for the $8$ WSC case. A second observation is that when the  average load is $20\%$ of the system capacity the communication complexity is constant, $0.987$, for both configurations, $4$ and $8$ WSCs, regardless of the choices of simulation parameters. At high average load, the same figure is confined to a small range, $4.078$ to $6.734$.

At low average load we also see a reduction of the average rejection ratio for both $4$ and $8$ WSCs,  $0.1 - 0.3$ (see Tables \ref{ResultsBidTab1} - \ref{ResultsBidTab3}). The results in Table \ref{ResultsBidTab4} show that increasing the distribution of the service time from the $ 1- 10$ to $1 - 20$ range increases the rejection rate; this is most likely due to the fact that we simulate only the execution over $200$ reservation slots and requests with a large service time arriving during lather slots do not have time to complete. At high average system load the average rejection ratio is only slightly better in case of market-based versus hierarchical control. Lastly, the market-based mechanism performs slightly better than hierarchical control in terms of slot-by-slot load balancing, the coefficient of variation of the system load per slot $\gamma \le 1.115$ 

The number of different service types offered by the cloud does not seem to affect the performance of the system and neither does the number of services supported by individual servers, as we can see from the results reported in Tables \ref{ResultsBidTab2}  and \ref{ResultsBidTab3}. The organization is  scalable, the results for $4$ and for $8$ WSCs differ only slightly. This is expected because of the distributed scheme where each WSC acts independently, it receives an equal fraction of the incoming  service requests and matches them to the bids placed by the servers it controls. 

\begin{figure*}[!ht]
\begin{center}
\includegraphics[width=8.7cm]{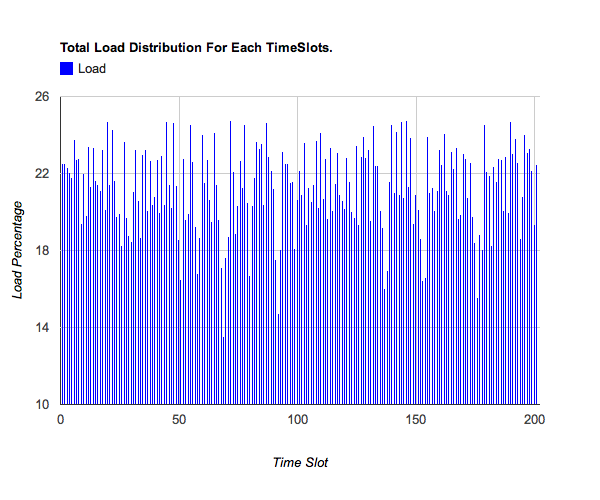}
\includegraphics[width=8.7cm]{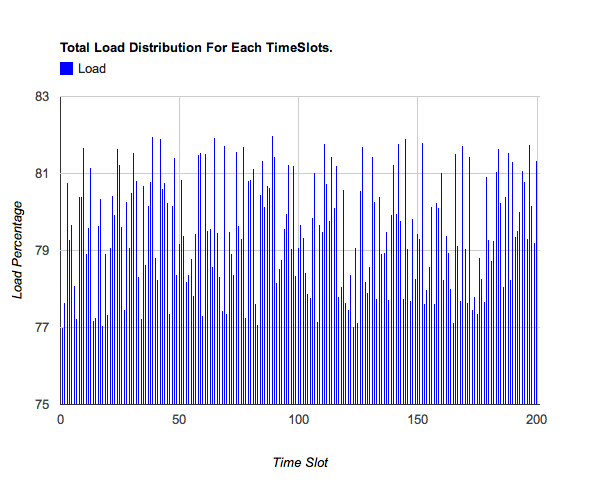}\\
\end{center}
\caption{Hierarchical control - time series of the average load  of a cloud with eight WSCs. The monitoring interval is $20$ reservation slots and the service time is uniformly distributed in the range $1 - 20$ reservation slots. The initial average system load is: (Left) $20\%$; (Right) $80\%$ of system capacity.}
\label{ResultsBid-Fig1}
\end{figure*}

\begin{figure*}[!ht]
\begin{center}
\includegraphics[width=8.7cm]{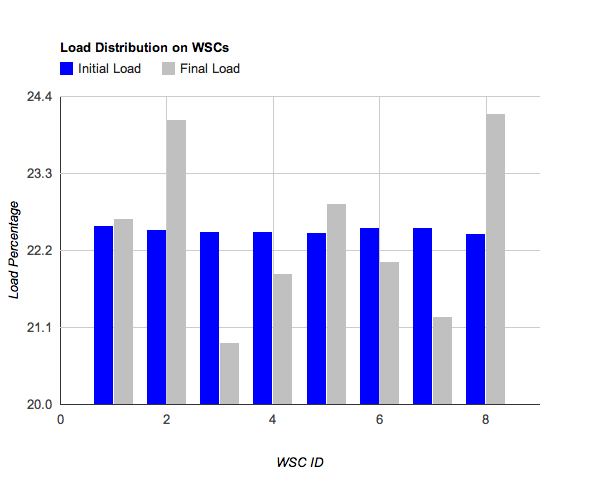}
\includegraphics[width=8.7cm]{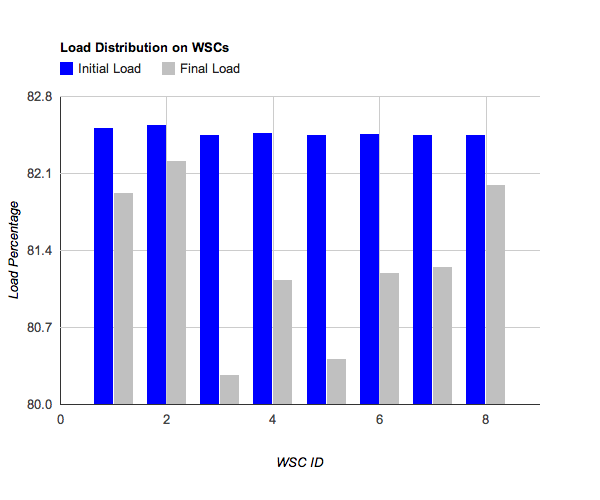}\\
\end{center}
\caption{Hierarchical control - initial and final average load of a cloud with eight WSCs. The monitoring interval is $20$ reservation slots and the service time is uniformly distributed in the range $1 - 20$ reservation slots. The initial average system load is: (Left) $20\%$; (Right) $80\%$ of system capacity.}
\label{ResultsBid-Fig2}
\end{figure*}

Figures \ref{ResultsBid-Fig1} and  \ref{ResultsBid-Fig2} refer to the case presented in Table \ref{ResultsBidTab4}  when the monitoring interval is $20$ time slots and the service time is uniformly distributed in the $(1 - 20)$ slots range and there are 8 WSCs.  Figures \ref{ResultsBid-Fig1}(Left) and (Right) show the time series of the average system load for the low and the high initial load, respectively.  The actual system workload has relatively  small variations from slot to slot; for example, at high load the range of the average system load ranges from $77\%$ to $82\%$ of the system capacity.  Figures  \ref{ResultsBid-Fig2} (Left) and (Right)  show the initial and the final load distribution for the 8 WSCs; the imbalance among  WSCs at the end of the simulation is in the range of $1-3\%$ at low load and between $80.1\%$ and $80\%$ at high load.

\section{Conclusions and Future Work}
\label{Conclusions}

Low average server utilization \cite{Snyder10} and its impact on the environment \cite{Blackburn10}, the increasing heterogeneity of cloud servers, and the diversity of services demanded by the cloud user community are some of the reasons why it is imperative to devise new resource management policies. These policies should significantly increase the average server utilization and the computational efficiency measured as the amount of computations per Watt of power, make cloud computing more appealing and lower the costs for the user community, and last, but not least, simplify the mechanisms for cloud resource management.

Individual Cloud Service Providers believe that they have a competitive advantage due to the uniqueness of the added value of their services \cite{Marinescu13} thus, are not motivated to disclose relevant information about the inner working of their systems. A rare glimpse at the architecture of the cloud is provided in \cite{Barroso13} and we have used this information to investigate a realistic model of the cloud infrastructure. 

We defined several high-level performance measures for the resource management mechanism of  a hierarchically organized cloud including the communication complexity of the decision making process, the ability to balance the load in time and respond to incoming service requests. Admittedly, there are several other performance measures such as the robustness of the mechanisms, the ability to accommodate system changes, and so on but a more complex model of the system would make simulation prohibitively costly and require a longer time to obtain the results.  

We choose to compare hierarchical control based on monitoring, a clear choice for a hierarchically organized system, with a bidding scheme whose appeal is its simplicity.  This simplicity has important consequences; for example, the admission control is implicit, when the system is heavily loaded there are no bids. Moreover, failed servers  are detected because they do not participate in the bidding process, and so on. We used simulation to carry out our investigation simply because it is unfeasible to evaluate analytical models and experiments with systems at this scale are not possible. Even so the simulations we conducted are computationally intensive and it took days to get the results of a single experiment.

It was not unexpected to see that the average system load affects the performance of  resource management mechanisms; the higher the load the larger the overhead due to communication, the lower the ability to balance the load, and higher the ratio of requests rejected. But our results show that  
the communication complexity is orders of magnitude lower for the economic model than that of hierarchical control. Our aim is a qualitative, rather than a quantitative analysis, and a discrepancy of orders of magnitude provides solid arguments for the advantages of one approach versus the other.

In some sense the hierarchic cloud resource management is scalable. Our results show that at low and at high load a system with $8$ WSCs has a similar behavior as the one with only 4 WSCs; at high load both configurations are affected by increased overhead. It is particularly significant the fact that at high load the number of messages as the cell level increases three fold. Typically,  computer clouds use 10 Gbps Ethernet networks; the contention for cell-level and WSC-level connectivity will further limit the ability of the hierarchical control model to perform under stress and lower considerably the bandwidth available for applications.

The economic cloud resource management model using bidding considerably more performant than the hierarchic control in virtually all aspects. The communication complexity is lower, it better balances the load among several WSCs, ensures a more even load distribution in time, requires less space to store the information required to perform the resource management, it is robust, and some of the policies such as admission control are implicit. The most significant result is that this model performs very well when the system is heavily loaded, when the average server load is around $80\%$ of its capacity. 

The simulation time is prohibitive and would dramatically increase for a heterogeneous cloud infrastructure model where individual servers provide one type  of service rather than several. Such a more realistic model is likely to make a hierarchical control model even less attractive.

The economic model discussed in this paper should be seen as an intermediate step towards the ideal case when self-organization and self-management  drive the cloud infrastructure. Self-organization is the spontaneous emergence of global coherence out of local interactions. A self-organizing system responds to changes in the environment through adaptation, anticipation, and robustness. The system reacts to changes in the environment, predicts changes and reorganizes itself to respond to them, or is robust enough to sustain a certain level of perturbations. Our future work is focused on cloud self-organization and self-management based on coalition formation and combinatorial auctions \cite{Marinescu14}.

\end{document}